\documentstyle{article}
\begin{document}
\begin{center}
{\Large \bf Ballistic electronic transport in Quantum Cables}
\end{center}

\vspace{.3cm}
\noindent
\begin{center}
{\large Z. Y. Zeng$^{\ddagger}$, Y. Xiang, and L. D. Zhang}
\end{center}

\vspace{.001cm}
\noindent
\begin{flushleft}
{\it Institute of Solid State Physics, Chinese Academy of
Sciences, \rm P.O. Box 1129, Hefei,\\ 230031, P. R. China \\}
\end{flushleft}

\vspace{.5cm}
\noindent
\begin{center}
{\bf Abstract}
\end{center}
We studied theoretically ballistic electronic transport in a proposed
mesoscopic structure - Quantum Cable. Our results demonstrated that
Qauntum Cable is a unique structure for the study of mesoscopic
transport. As a function of Fermi energy, Ballistic conductance
exhibits interesting stepwise features. Besides the steps of one or two
quantum conductance units ($2e^2/h$), conductance
plateaus of more than two quantum conductance units can also be expected
due to the accidental degeneracies (crossings) of subbands.
As  structure parameters is varied,  conductance width
displays oscillatory properties arising from the inhomogeneous
variation of energy difference betweeen adjoining transverse
subbands.
In the weak coupling limits, conductance steps of height $2e^2/h$
becomes the first and second plateaus for the Quantum Cable of
two cylinder wires with the same width.

\vspace{1.cm}

\newpage

\noindent

\begin{center}
{\bf I. INTRODUCTION}
\end{center}

Early investigations of ballistic electronic transport through
microconstrictions led to the discovery of
conductance quantization. Initially this
phenomenon was observed in two-dimensional electron gas (2DEG) systems,
 manifesting
itself in $2e^2/h$ steplike variations of the conductance as a function of
the transverse size of the narrowing$^1$. It originates from the discrete
character of propagating modes through the constriction due to the
qauntization of the transverse momentum.  Under approriate conditions,
this phenomenon of conductance quantization should also occur in 3D
point contacts with small constriction diameters$^2$.
Ballistic electron
transport through narrow constrictions of different confining potential$^3$
has been extensively studied based on Landauer and B{\" u}ttiker
approach$^4$.
For a long constriction (quantum wire or quantum waveguide), ballistic
conductance is directly propotional to the interger number of propagating
modes or conductance channels and increases with the constriction width or
Fermi energy in steps of $2e^2/h$  each time a new channel opens up.

 Owing to the recent advances in modern nanotechnologies, such as
  molecular-beam epitaxy,
  electron-beam lithography, self-assembled growth etc. ,
 it has been possible to design and fabricate various kinds of mesoscopic
 devices$^5$, in which electron can retain phase coherence
 while traveling through the active region. Successful experimental
 demonstration of mesoscopic devices have stimulated in turn much theoretical
 interest in exploiting the wave nature of electrons$^{5,6}$.
 Mesoscopic devices
 are different from
 traditional electron devices in principle. In analyzing and
 designing mesoscopic quantum devices,  it is essential to take into
 account the quantum charater of electrons.
 Many kinds of quantum devices with various functions have been proposed
 and  designed$^{5,6}$. Research on these mesoscopic structures revealed many
 phenomena such as the quantized conductance of point contacts, persistent
 current through mesoscopic metallic ring, universal conductance fluctuations,
 Coulomb blockade, resonant tunneling, etc.$^7$ In particular, propagation of
 electrons along quantum wires of various geometries has been considered
 extensively$^8$. More recently, del Alamo and Eugster have proposed and
 fabricated a new kind of mesoscopic device-coupled dual 2D quantum wires
 (2D CDQW)
 structure as field-effect directional couplers$^9$. The operation principle
 of this device is based on the tunneling effects between electron
 waves propagating in the two adjacent waveguides through a controllable
 potential barrier. Later some groups$^{10}$
 investigated its ballictic electronic
 transport properties  with and without the application of magnetic field.

    A new kind of nanostructure referred to as Coaxial Nanocable has
    been succesfully
 synthesized by Suenaga et al.$^{11}$ in 1997 and by Zhang et al.$^{12}$
  in 1998. It
 comprises a solid and a hollow conducting cylinders of mesoscopic
 size seperated by a
 insulating layer. If this insulating layer is not thich enough to
 forbiden electrons' tunneling, it can be viewed as a coupled mesoscopic
 structure formed by two 3D quantum wires and a cylindrical potential barrier.
 This inspired us to propose a similar mesoscopic structure,
 Quantum Cable$^{13}$,
 in which electron are confined in a Nanocable-like composite potential
 wells; i.e., electrons in the inner wire are subjected to a solid
 cylinder potential well, while electrons in the outer wire are limited
 into a hollow cylinder well; and they can tunnel into each other wire
 through the cylindrical coupling barrier. The structure of Quantum Cable is
 schematically shown in Fig. 1. Our previous calculations$^{13}$
  showed that Quantum
 Cable has perticular energy subband spectrum different from the 2D CDQW.
 For ballistic conductance, it is believed that there exists some significant
 discrepancies between the 2D and 3D quantum wire$^{14}$. Therefore we expect
 Quantum Cable will exhibit unique transport properties, which are unexpected
 in the usual 2D CDQW structures.

   The paper is organized as follows. In Section II, we present the
   formulas for calculating the ballistic conductance from Landauer and
   B{\" u}tticker's formula. Section III gives numerical calculations
   for ballistic conductances in some cases. We summarize
   our results in Section IV.

\begin{center}
{\bf II. MODEL AND FORMULATION}
\end{center}

  We consider ballistic electronic transport through the Qauntum Cable
  structure between two large bulk reservoirs.
  The Qunatum Cable comprises two coaxial cylindrical quantum
  wires - the inner wire is a solid cylinder well with radius $R_1$
  and the outer a hollow cylinder well with inner radius $R_2$ and
  outer radius $R_3$.  Two cylindrical wires
   are coupled through a thin layer of
  potential barrier of width $R_B=R_2-R_1$ and height $U_B$.
  The electrons are free to move along the $z$ axis of Quantum Cable ,
   whereas their  motion in the radial direction is quantized.
   The cable length $L$ is assumed to be, on the one hand, small compared to
   the elastic and inelastic mean free paths of the electrons ( ballistic
   limit), and, on the other hand, large compared to the normal
   coherence length to ensure the absence of backscattering effects.
   Then the ballictic conductance of the Quantum Cable, $G$, is determined
   by the Landauer-B{\" u}ttiker formula$^4$
   \begin{equation}
   G=\frac{2e^2}{h} \sum T_{nl;n'l'},
   \end {equation}
   where $T_{nl;nl''}$ is the transmission probability of electrons
   from the incident transverse mode $(n,l)$ to the outgoing mode
   $(n',l')$. The sum extends over all possible channels. If the applied
   contacts is adiabatic (i.e., whose dimensions changes slowly on the
   scale of the Fermi wavelength), and electron scattering in the
   contact regions is very weak, then the mode mixing can be neglected
   ($T_{nl, n'l'}=0$  if  $n\not = n'$ or $l\not = l'$)
   and the transmission into the same channel
   $T_{nl,nl}=1$. This allows us to express the conductance at low
   temperature as
   \begin{equation}
   G=\frac{2e^2}{h} \sum_{n,l} \Theta(E_F-E_{nl}),
   \end{equation}
   where $E_F$ is the Fermi energy and $E_{nl}$ is the transverse part of the
   electron enenrgy in Quantum Cable, which satisfies the follow relation
   derived from the standard effective-mass boundary conditions
\begin{eqnarray}
 F_1(k_2,k_3;R_1,R_2,R_3)+\frac{m^*_3k_2}{m^*_2k_3}
                 F_2(k_2,k_3;R_1,R_2,R_3)+
\frac{m^*_2k_1}{m^*_1k_2}\frac{J'_n(k_1R_1)}{J_n(k_1R_1)} \times \nonumber \\
{[G_1(k_2,k_3;R_1,R_2,R_3)+\frac{m^*_3k_2}{m^*_2k_3}G_2(k_2,k_3;R_1,R_2,R_3)]}=0,
\end{eqnarray}
where
\begin{eqnarray}
 F_1(k_2,k_3;R_1,R_2,R_3) & = &  [K_n(k_2R_2)I'_n(k_2R_1)-K'_n(k_2R_1)
                                   I_n(k_2R_2)] \times  \nonumber \\
                          &   &  [J_n(k_3R_3)Y'_n(k_3R_2)-J'_n(k_3R_2)
                                     Y_n(k_3R_3)],  \nonumber \\
 F_2(k_2,k_3;R_1,R_2,R_3) & = & [K'_n(k_2R_1)I'_n(k_2R_2)-K'_n(k_2R_1)
                                I'_n(k_2R_2)] \times  \nonumber \\
                          &   & [J_n(k_3R_3)Y_n(k_3R_2)-J_n(k_3R_2)
                                      Y_n(k_3R_3)], \nonumber \\
 G_1(k_2,k_3;R_1,R_2,R_3) & = &  [K_n(k_2R_1)I_n(k_2R_2)-K_n(k_2R_2)
                                 I_n(k_2R_1)] \times \nonumber \\
                          &   &  [J_n(k_3R_3)Y'_n(k_3R_2)-J'_n(k_3R_2)
                                  Y_n(k_3R_3)], \nonumber \\
 G_2(k_2,k_3;R_1,R_2,R_3) & = &    [K'_n(k_2R_2)I_n(k_2R_1)-K_n(k_2R_1)
                                    I'_n(k_2R_2)] \times \nonumber \\
                          &   &  [J_n(k_3R_3)Y_n(k_3R_2)-J_n(k_3R_2)
                                   Y_n(k_3R_3)],
 \end{eqnarray}
 where $ k_1 = [(2m^*_1/\hbar^2)E_{nl}]^{1/2}, k_2 =
 [(2m^*_2/\hbar^2)(U_B-E_{nl})]^{1/2}, k_3 = [(2m^*_3/\hbar^2)E_{nl}]^{1/2}$;
 $m^*_i (i=1,2,3)$ is the electron effective mass;
and, $J_n$ is the Bessel function of first kind,
  $Y_n$ the Bessel function of second kind and $K_n, I_n$ are the modified
  Bessel functions$^{15}$, respectively;
  $f'(x)=df(x)/dx$.
  In the course of deriving Eq. (3), we adopted the hard - wall model, i.e.,
  $U(\rho)=0$ for $\rho \geq R_3$. From the definitions and properties of
   Bessel  functions$^{15}$,
  it can be deduced readily that, eigenstates with nonzero azimuthal
  quantum number are doubly degenerate ($E_{nl}=E_{-nl}$), which is the
  result of cylindrical symmetry of Quantum Cable.
  Combining Eqn. (2) with Eqn. (3), one can calculate the ballistic
  conductances in an accurate way.  Electron's
  effective masses in different layers, in our calculations, are set as
   $m^*_1=5.73\times 10^{-32} kg, m^*_2=1.4 m^*_1$
  $m^*_3=m^*_1$, as one did in the usual $GaAs/Ga_{0.7}Al_{0.3}As$
  heterostructures.

\begin{center}
{\bf III. RESULTS AND DISCUSSION}
\end{center}

  For comparison, we begin with the study of ballistic conductances
  versus Fermi energy
  of single solid and hollow quantum cylinders and single
  2D quantum waveguide. In Fig. 1 we give the ballistic conductances for
  2D quantum waveguide, solid quantum cylinder and hollow quantum cylinder
  and their corresponding subband dispersions.
   It follows from Eqn. (2)
  that the ballistic conductance rises in step-like manner with the
  increasing Fermi energy, and that the distribution of the conductance
  stpes is determined by the transverse mode energy $E_{nl}$. For 2D quantum
  waveguide structure, electronic states in the channel
  are characterized by only one quantum number $n$,
  and, the transverse energy is given by
  $E_n=n^2\pi^2\hbar^2/(2m^*_1a^2) (n=1, 2, 3, \cdots)$, where $a$ is the
  width of the waveguide. Then conductance would increase a quantum
  conductance unit $2e^2/h$ each time
  Fermi energy goes across  a transverse subband; the step
  width increases with the increment of Fermi energy and is decided by
  the energy difference between neighboring subbands. The step-fashioned
  conductance  vs Fermi energy was plotted in Fig. 1 (a) for single
  2D quantum waveguide of width $210$ nm.  In the cases of
  3D cylindrical quantum wire structure,  because of the azimuthal
  degeneration  of the transverse modes (electrons with azimuthal
  quantum number $n$ and $-n$ have the same transverse momenta) steps of
  one ($n=0$) or two quantum conductance units ($n\not=0$)
   can be observed. While the
  steps of two quantum conductance units can only be expected in the
  weekly-coupling two 2D quantum waveguides at low Fermi energies due to
  the twofold degeneracy of lowest subband energy dispersions.
  To verify this,
  we plotted the ballistic conductance for single solid quantum cylinder
  of radius $60$ nm in Fig. 1 (b),
   and that for single hollow quantum cylinder
  of inner radius $100$ nm and outer radius $150$ nm in Fig. 1 (c).
  It is evident that, steps of one or two quantum conductance units appeared.
  We also noticed that nonuniform distribution of conductance step width for
  solid cylinder; similar plateau-like
   structure as the 2D waveguide exists in the
  ballistic conductance of hollow cylinder within consecutive energy domains.
  This feature can be understood from the subband spectrum of hollow
  cylinder,
  since its energy difference between transverse subbands
  belong to the same azimuthal quantum number $n$
  will be increased with the increase of the radial quantum number $l$
  in the similar way as 2D quantum waveguide's, while
  that between states of
   the different azimuthal quantum number $n$ is comparatively large,
   thus one can
   observed its similar step-like conductance structure within consecutive
   energy regions
   as that in 2D waveguide case.

   As a solid quantum cylinder is coupled with a hollow cylinder through a
   tunable barrier, Quantum Cable structrue is formed. It is therefore
   expected that coupling effects would reflected in the ballistic
   conductance profile. Fig. 3 presents the calculation results of
   ballistic conductance vs Fermi energy of Quantum Cable with
   structure parameters $R_1=30$ nm, $R_3-R_2=30$ nm, $U_B=20$ mev
    for variational
   barrier width $R_B$.  With the increase of barrier width, that is to say,
   as the coupling between two cylinder wires becomes weak, more narrower
   conductance plateaus are displayed, some of conductance plateaus
   tend to be narrower while some becomes broader; some other
   becomes narrower first and then broader. This individuality origins also
   from the unique subband structure of Quantum Cable,
    energy spacing between consecutive subbands varies
   inhomogeneously with the broadenness of the coupling barrier,
   as shown in our previous work$^{13}$. In the
   extreme limit (i.e., $R_B \rightarrow \infty$ or
   $U_B \rightarrow \infty$), Quantum Cable
   turns into the uncoupled cylinders structure. Then ballistic
   conductance spectrum is simplily determined by the crude arrangement
   of subbands of solid and hollow quantum cylinders, and will show
   conductance charaters of both solid and hollow cylindrical wires, which
   can be easily found in the $R_B=20$ nm case. Another observable
   phenomenon is the gathering of conductance steps within Fermi
   energy $4-7$ mev due to the subband bundling effects$^{13}$ for the
   weakening of the coupling between two cylinders. It should be
   noticed that, steps of three and four quantum conductance units
   can also be expected to be observed (see, for example, $R_B=8$ nm case and
   $U_B=35$ mev case in Fig. 4).
   This feature stems from the accidental degeneracies (crossings)
    of transverse modes caused by some kind of symmetry.
   Since the subband (0,0) keeps the ground subband albeit of
   Cable structure parameters, and the subband (0,1) tends to be
   the first excited subband as the coupling becomes weak enough,
   thus one can find that the height of first conductance step is always
   $2e^2/h$, and that of the second will be also one quantum conductance
   unit as the coupling is weak enough; in other words, the steps of
   one quantum conductance unit tend toward low Fermi energy region
   as the barrier width increases.

        An alternative way to change the coupling strength between two
        wires is that the variation of coupling barrier height. In Fig. 4
        we give the conductances for different barrier height from
        $0$ to $40$ mev. As in the case of increased barrier width,
        ballistic conductances for lifted barrier height
        show similar features, except that, the total number of
        conductance plateaus  increases in the former case while it
        is reduced in the later case. The reason is that the energies of
        most subbands tend to decrease for increasing width but increase
        for raising barrier height, as shown in the Cable subband spectrum
        in both cases$^{13}$.

    In general, conductance plateaus of very narrow width or steps
    of two quantum conductance units are  expected for
    the 2D weakly coupled quantum waveguides with two same width waveguides,
    and they will be violated by the deviation of the width of one of the
    waveguides. Since, for such structure, crossing occurs between two
    states with the same symmetry$^{16}$,
     then if the widths of the two waveguides
    are not the same, its defining potential will be clearly asymmetric
    and thus its eigensubbands do not exhibit subband crossings. However,
    plateaus of narrower width or of three or four quantum conductance
    units  can be observed in the Quantum Cable of cylinders with different
    radius. For Quantum Cable structure, subband crossings is foreign to the
    parameters associated with cylinders radia. In Fig. 5, conductances
    as a function of Fermi energy  are plotted for
    different outer cylinder radius $R_1$, where $R_B=5$ nm, $R_3-R_2=
    30$ nm, $U_B= 20$ mev. Although one does not see the
    steps of three or four conductance units, they can be assured to
    exist in the case of $R_1$ being near $31$ nm from the conductance
    curves for $R_1=30, 31, 32$ nm, or directly from the relation of
    subband energy with the inner cylinder radius$^{13}$. In addition, one
    can notice that the second step of only one quantum conductance unit
    may not always move to abut against the first step of one conductance
    unit. We then can conclude that conductance steps of one quantum
    conductance unit $2e^2/h$ appear as the first and second steps only if the
    cylindrical wires of Quantum Cable are of the same width.

\begin{center}
{\bf IV. CONCLUSIONS}
\end{center}

In this paper, we studied ballistic electronic transport in Quantum Cable
formed by a solid cylinder, hollow cylinder and a cylindrical coupling
potential barrier. Because of the azimuthal degeneration of the
transverse modes (electrons with azimuthal quantum number $n$ and $-n$ have
the same transverse momenta), steps of one or two quantum conductance units
can be observed in solid and hollow quantum cylinders;
 while in 2D quantum
waveguide, steps of one quantum conductance unit can only be expected.
Since Quantum Cable is constructed by a solid and a hollow quantum cylinder
which are coupled by a tunable potential barrier, apart from the
above features, conductance will displays some properties originated from
the coupling effects. As one of two cylinder widths or
the width or height of the coupling barrier is
varied, conductance plateaus of more than two quantum conductance units
could be seen, due to the fact that transverse modes will cross for
some parameters. Arising from the inhomogeneous variation of
energy difference between adjoining subbands, some ballistic conductance
plateaus exhibit oscillation. One plateau with conductance height of
$2e^2/h$ is always the first conductance step irrespective of parameter
values,  since the transverse mode $(0,0)$ keeps the ground subband whatever
the value of structure parameters. As the coupling between two cylinders
of the same width becomes weak, i.e.,
 the width or height of barrier increases, the other
step of only one quantum conductance unit $2e^2/h$ shifts into the second
ballistic conductance step. This phenomenon can also be explained from
the subband spectrum of Quantum Cable.

\begin{center}
{\bf ACKNOLEDGEMENT}
\end{center}

  This work is supported by a key project for fundamental research
in the National Climbing Program of China.

\vspace{.5cm}
\noindent
{\bf References}

\vspace{.1cm}
\noindent
\hspace{0.2cm} $\ddagger$ E-mail address: zyzeng@mail.issp.ac.cn
\begin{enumerate}
\item H. Van Houten, C. W. J. Beenakker, and B. J. von Wees, Semicond.
Semimet. {\bf 35}, 9 (1992).
\item J. A. Torres, J. I. Pascual, and J. J. Saenz, Phys. Rev. B
{\bf 49}, 16581 (1994); A. G. Scherbakov,
E. N. Bogachek, and Uzi Landman, ibid.
 {\bf 53}, 4054 (1996).
\item G. Kirczenow, Solid State Commun. {\bf 68},715 (1988);
A. Szafer and A. D. Stone, Phys. REv. Lett. {\bf 62}, 300 (1989);
N. Garcia and L. Escapa, Appl. Phys. Lett. {\bf 54}, 1418 (1989);
E. G. Haanappel and D. van der Marel, Phys. Rev. B {\bf 39} 5484 (1989);
A. Matulis and D. Segzda, J. Phys. Condens. Matter {\bf 1}, 2289 (1989);
M. Yosefin and M. Kaveh, Phys. Rev. Lett. {\bf 64}, 2819 (1990);
E. Tekman and S. Ciraci, Phys. Rev. B {\bf 43}, 7145 (1991).
\item R. Landauer, Philos. Mag. {\bf 21}, 863 (1970);
M. B{\" u}ttiker, Phys. Rev. Lett. {\bf 57}, 1761 (1986).
\item S. Datta and M. J. McLennan, Rep. Prog. Phys. {\bf 53}, 1003 (1990);
T. J. Thornton, ibid. {\bf 57}, 311 (1994).
\item F. Sols, M. Macucci, U. Ravaioli, and K. Hess,
Appl. Phys. Lett. {\bf 54}, 350 (1989); Y. Avishai and Y. B. Band,
Phys. Rev. B {\bf 41}, 3253 (1990); J. Wang, Y. J. Wang, and H. Guo,
ibid. {\bf 46}, 2420 (1992); Z. -L. Ji and K. -F. Berggren,
ibid. {\bf 45}, 4662 (1992);
H. Xu,ibid. {\bf 47}, 9537 (1993)
 H. U. Baranger and P. A. Mello,
Phys. Rev. Lett. {\bf73}, 142 (1994).
\item Mesoscopic Phenomena in Solids,edited by B. L. Altshuler, P.A. Lee and
R. A. Webb, (Elsevier Science Publishers B. V. ) (1991).
\item S. Luryi and F. Capasso, Appl. Phys. Lett. {\bf 47}, 1347 (1985);
S. Datta, M. R. Melloch, S. Bandyopadhyay, R. Noren, M. Vaziri,
M. Miller, and R. Reifenberger, Phys. Rev. Lett. {\bf 55}, 2344 (1985);
T. J. Thornton, M. Pepper, H. Ahmed, D. Andrews, and
G. J. Davies, Phys. REv. Lett. {\bf 56}, 1198 (1986);
K. Tsubaki and Y. Tokura, Appl. Phys. Lett. {\bf 53}, 859 (1989);
R. Q. Yang and J. M. Xu, Phys. Rev. B {\bf 43}, 1699 (1991).
\item
J. A. del Alamo and C. C. Eugster, Appl. Phys. Lett. {\bf 56}, 78 (1990);
Phys. Rev. Lett. {\bf 67}, 3586 (1991).
\item C. C. Eugster , J. A. del Alamo, M. J. Rooks, and M. R. Melloch,
Appl. Phys. Lett. {\bf 60}, 642 (1992);
C. C. Eugster, J. A.  del Alamo, M. R. Melloch, and M. J. Rooks,
Phys. Rev. B {\bf 46}, 10406 (1992);
ibid {\bf 48}, 15057 (1993);
J. Wang, H. Guo,  and R. Harris, Appl. Phys. Lett. {\bf 59}, 3075 (1991);
J. Wang, Y. J. Wang, and R. Harris, Phys. Rev. B {\bf 46}, 2420 (1992);
J. R. Shi and B. Y.  Gu, ibid. {\bf 55}, 9941 (1997).

\item X. Suenaga, C. Colliex,  N. Demoncy, A. Loiseau, H. Pascard,  and
F. Willaime, Science {\bf 278}, 653 (1997).
\item Y. Zhang, K. Suenaga, C. Colliex and S. Iijima,   Science {\bf 281},
 973 (1998).
\item  Z. Y. Zeng, Y. Xiang, and L. D. Zhang (unpublished).
\item E. N. Bogachek, A. N. Zagoskin, and I. O. Kukik, Fiz. Nizk. Temp.
{\bf 16}, 1404 (1990) [Sov. J. Low. Temp. Phys. {\bf 16}, 796 (1990)].
\item Handbook of Mathematical Functions, edited by M. Abramowitz and
I. A. Stegun (New York, Dover) (1972).
\item  M. A. Morrison, T. L. Estle, and N. F. Lane,
Quantum States of Atoms, Molecules, and Solids (Prentice - Hall Inc.,
New Jersey) (1976).

\end{enumerate}

\newpage
\noindent
\begin{center}
{\bf Figure Captions}
\end{center}
\vspace{0.5cm}
\noindent
{\bf Fig.~1}~ A schematic view of ballistic Quantum Cable structure, where
$R_1$ is the radius of inner solid quantum cylinder, $R_2$ and $R_3$ are
the inner radius and outer radius of outer hollow cylinder, respectively;
two cylinders are coupled through a tunable potential barrier of width
$R_B=R_2-R_1$ and height $U_0$.

\vspace{0.5cm}
\noindent
{\bf Fig.~2}~   Conductance as a function of Fermi energy
and the corresponding energy dispersion for (a) 2D quantum waveguide of
width $210$ nm, (b) solid quantum cylinder of radius $60$ nm and (c)
hollow quantum cylinder of inner radius $R_2=100$ nm and outer radius
$R_3=150$ nm.

\vspace{0.5cm}
\noindent
{\bf Fig.~3}~ Conductance of Quantum Cable
as a function of Fermi energy for different barrier width $R_B$, where
structure parameters are chosen such that $R_1=30$ nm, $R_3-R_2=30$ nm,
$U_B=20$ mev.

\vspace{0.5cm}
\noindent
{\bf Fig.~4}~ Conductance of Quantum Cable
as a function of Fermi energy for different barrier height  $U_B$, where
structure parameters are chosen such that $R_1=30$ nm, $R_3-R_2=30$ nm,
$R_B=5$ mev.

\vspace{0.5cm}
\noindent
{\bf Fig.~5}~ Conductance of Quantum Cable
as a function of Fermi energy for different solid cylinder radius
$R_1$, where
structure parameters are chosen such that $R_B=5$ nm, $R_3-R_2=30$ nm,
$U_B=20$ mev.

\end {document}